\documentclass[preprint,12pt]{elsarticle}

\usepackage{amssymb}
\usepackage{amsmath}

\journal{Acta Astronautica}

\begin{document}

\begin{frontmatter}



\title{On the likelihood of non-terrestrial artifacts in the Solar System}


\author[bms,psu1]{Jacob Haqq-Misra\corref{cor1}}
\ead{haqqmisra@psu.edu}

\author[psu2]{Ravi Kumar Kopparapu}
\ead{ravi@gravity.psu.edu}

\address[bms]{Blue Marble Space Institute of Science, PO Box 85561, Seattle, WA 98145, USA}
\address[psu1]{Rock Ethics Institute, Pennsylvania State University, 201 Willard Building, University Park, PA 16802, USA}
\address[psu2]{Department of Geosciences, Pennsylvania State University, 443 Deike Building, University Park, PA 16802, USA}

\cortext[cor1]{Corresponding author}

\begin{abstract}
Extraterrestrial technology may exist in the Solar System without our knowledge. This is because the vastness of space, combined with our limited searches to date, implies that any remote unpiloted exploratory probes of extraterrestrial origin would likely remain unnoticed. Here we develop a probabilistic approach to quantify our certainty (or uncertainty) of the existence of such technology in the Solar System. We discuss some possible
strategies for improving this uncertainty that include analysis of moon- and Mars-orbiting satellite data as well as continued exploration of the Solar System.
\end{abstract}


\end{frontmatter}


\section{Introduction}

If extraterrestrials exist in our galaxy, then might they try observing
us? This question was raised by Ronald Bracewell \cite{bracewell1960} shortly after
Cocconi and Morrison's suggestion \cite{cocconi1959} to search for extraterrestrial
broadcasts. It is certainly possible that extraterrestrials could
be observing us remotely; after all, near-future human technology
includes the prospects of observing atmospheric spectra of extrasolar
terrestrial planets \cite{desmarais2002} and exploring nearby star systems
with unpiloted remote exploratory probes \cite{matloff2000,bjork2007,cotta2009}. One possible scenario for human exploration
of space begins with the discovery of an Earth-like planet around
a nearby star, with an exploratory probe sent as a follow-up mission.
It is at least plausible that extraterrestrials might also adopt a
similar strategy \cite{freitas1983,cornet2003}. If so, then
extraterrestrial technology could be hiding in our own Solar System.

We have obviously not yet discovered any technology of extraterrestrial
origin, but how sure can we be that the Solar System contains none
of these artifacts? In this paper we develop a probabilistic approach
to this question. Certain regions of space have been searched sufficiently
to discover such technology, if it exists, but much of the Solar System
remains unexplored. It is completely possible that extraterrestrial
technology could be actively observing us without our knowledge, while
it is also possible that defunct extraterrestrial technology exists
on planetary surfaces or interplanetary space. Therefore, it is of
interest to calculate a likelihood that a given region of
space is in fact absent of extraterrestrial technology. We begin our
argument by first clarifying the types of extraterrestrial technology
we consider. We then develop our probabilistic analysis and discuss
its implications for future searches of the Solar System.

\section{The Testable Zoo Hypothesis}

The Fermi paradox \cite{hart1975}, also known as the Great Silence \cite{brin1983}, asks the question: if intelligent life is common in the galaxy,
then why have no technological civilizations been observed yet? Many
resolutions have been proposed to this question \cite{webb2002,cirkovic2009}. These solutions include the possibility that life is rare in
the galaxy \cite{ward2000} or that intelligent civilizations
inevitably destroy themselves \cite{shklovskii1966,kaplan1971,sagan1973}. However, the absence of extraterrestrials could simply
be because long-lived technological civilizations expand too slowly
to have yet arrived at Earth \cite{vonhoerner1975,newman1981,haqqmisra2009}. Still other resolutions to the Fermi paradox
suppose that extraterrestrials are actually widespread but have purposefully
refrained from making contact \cite{ball1973,deardorff1987,fogg1987,baxter2001}. In both of these cases, there is nothing to preclude
the possibility that extraterrestrials might remotely explore our
Solar System using unpiloted probes. We refer to such technology as
non-terrestrial Artifacts (NTAs) to indicate their extraterrestrial
origin but also to distinguish realistic remote exploratory probes
from piloted flying saucers in the realm of science fiction.

The \emph{zoo hypothesis} \cite{ball1973} is a term describing the possibility
that extraterrestrials might be concealing themselves from us for
one reason or another. We often imagine that a benevolent extraterrestrial
civilization might have our best interests in mind and avoid interfering
with our primitive culture until we cross a certain threshold, such
as the discovery of light-speed travel \cite{ball1973,deardorff1987}
or our initiation of conversation with a nearby intelligent space
probe \cite{tough1998}. An even more omnipotent extraterrestrial civilization
might engineer a virtual planetarium that forces us to observe an
empty universe \cite{baxter2001}, which would give us almost no chance
at discovering their presence until they choose to reveal themselves
to us. More mundane versions of the zoo hypothesis simply invoke the
vastness of space to suggest that extraterrestrials or extraterrestrial
technology could be hiding in the asteroid belt \cite{papagiannis1978,kecskes1998,kecskes2002,forgan2011b} or other places in the Solar System \cite{arkhipov1996,arkhipov1998, burkeward2000} without us noticing. This range of possibilities
for the zoo hypothesis can be rather confusing for SETI (search for extraterrestrial intelligence) researchers
because some of these hypotheses are testable whereas others are not.
We therefore sub-divide this zoo hypothesis into two categories: the
\emph{testable zoo hypothesis} and the \emph{untestable zoo hypothesis}.

The untestable zoo hypothesis describes any attempt by an advanced
extraterrestrial civilization to willfully conceal its presence from
us. We typically assume that any extant extraterrestrial civilization
is probably much more technologically advanced than us \cite{bracewell1960,cocconi1959,shklovskii1966,kaplan1971,sagan1973,tarter2001,baum2011}, so there could 
be almost no limit to the ways in which superior extraterrestrials could keep
us in the dark. In this sense, the untestable zoo hypothesis is tantamount
to an unknowable faith-based system because we can ascribe to the
extraterrestrials any mechanism imaginable that maintains their unobservability.
Such scenarios are popular in science fiction, perhaps in part because
they allow us to maintain belief in a Galactic Club \cite{bracewell1975}
of extraterrestrial civilizations that humanity might one day join.
However, these solutions to the Fermi paradox are completely untestable
and therefore a distraction to the scientific SETI discussion. (Recent 
calculations have even shown that the total hegemony supposedly 
required for a Galactic Club to maintain a zoo hypothesis policy would 
break down on a galaxy-wide scale \cite{forgan2011}.) We
cannot rule out the possibility that the untestable zoo hypothesis
accounts for the absence of extraterrestrials, but any further quantitative
analysis requires that we set aside this possibility and limit our
consideration to only the testable zoo hypothesis.

It is within the realm of possibility for human civilization to send
exploratory probes to other star systems, and a good deal of thought
has been already given to the associated engineering challenges \cite{matloff2000}. It is therefore plausible that extraterrestrials might attain
similar technology and remotely explore nearby planetary systems.
We refer to this set of possibilities as the testable zoo hypothesis
because in this case there is no willful attempt at concealment. It
is unlikely that our search efforts to date would have discovered
any NTAs in the galaxy because such interstellar probes would have
a limiting size of about 1 to 10 meters \cite{freitas1983}. (Objects with a much larger scale of 1000 km would be comparable in size to a minor planet and exert appreciable gravitational force on its surroundings, while the energy requirements to transport such a vehicle may be impermissible for interstellar travel. A smaller probe could implement a large lightweight solar sail in order to travel large distances using radiation pressure, but a massive solar sail would be difficult to keep concealed.) Thus, extraterrestrial
artifacts may exist in the Solar System without our knowledge simply
because we have not yet searched sufficiently. As we continue to explore
at improved spatial resolution, though, we will increase our chances
of discovering NTAs, if they exist, or else we will increase our confidence
that NTAs are absent. In either case, this testable zoo hypothesis
describes possibilities that can be examined through observation and
exploration, even if we believe the probability of finding a probe
to be low. In the next section, we develop a probabilistic framework
for estimating the probability that the Solar System is in fact absent
of NTAs.

Extraterrestrial visitors to the Solar System could also leave signatures 
other than technological artifacts. For example, a civilization could establish residence 
in the asteroid belt and take advantage of the abundance of mineral resources \cite{papagiannis1978,kecskes1998,kecskes2002,forgan2011b}.
Evidence of this type of visitation on asteroids or other rocky bodies could be inferred 
from the discovery of non-terrestrial mining technology or industrial byproducts such as
nuclear waste. Alternatively, extraterrestrial life could be so different from our own biology 
that it has escaped our notice; perhaps extraterrestrial microbes even exist today on Earth in a 
shadow biosphere parallel to our own \cite{davies2009,davies2010}! Signatures of extraterrestrials 
such as these are certainly interesting and worth keeping in mind as Solar System exploration 
progresses. However, non-technological signatures require that extraterrestrials have physically 
visited the Solar System, while the technological signature of an NTA could come from a distant 
extraterrestrial civilization that has only gazed at Earth from afar. Although we cannot preclude the 
possibility that extraterrestrials have visited the Solar System, for the purposes of this analysis we will 
limit our consideration to signatures such as NTAs that could arise from remote exploration.

\section{Probabilistic Analysis}

So far no search of the Solar System has found unequivocal evidence
of NTAs, but few, if any, of these attempts would be capable of detecting
a 1 to 10 meter probe. If NTAs do exist in near-Earth orbit, on the
moon, or other nearby locations, then they probably still remain undetectable
by humans. Because we cannot know if NTAs exist until we actually
detect one, it is therefore of interest to determine how sure we can be
that we should have already found any NTAs lurking in the Solar System. Or, to put it another
way: based on our searches to-date, how confident are we that the
Solar System is free of NTAs?

We approach this problem by considering a search for NTAs within a
volume $V$. We define the hypothesis $H$ = {}``NTAs are present
within a volume $V$'' so that its logical complement is $\bar{H}$
= {}``NTAs are not present within a volume $V$''. We also represent
incomplete and partial searches of $V$ with the sub-sampled volume
$\bar{V}_{R}$ = {}``A sub-sampling of $V$ at spatial resolution $R$''
that finds no NTAs. Using these statements, we can describe the conditional
probability $p(\bar{H}|\bar{V}_{R})$ as the probability that no NTAs exist
within $V$ given a null search $\bar{V}_{R}$ of any geometry. Following \cite{ambaum},
we can expand this probability using Bayes' theorem:

\begin{eqnarray}
p(\bar{H}|\bar{V}_{R}) & = & 1-p(H|\bar{V}_{R})\\
 & = & 1-p(\bar{V}_{R}|H)\frac{p(H)}{p(\bar{V}_{R})}\\
 & = & 1-p(\bar{V}_{R}|H)\frac{p(H)}{p(\bar{V}_{R}|\bar{H})p(\bar{H})+p(\bar{V}_{R}|H)p(H)}\\
 & = & 1-\frac{p(\bar{V}_{R}|H)}{\Theta(H)p(\bar{V}_{R}|\bar{H})+p(\bar{V}_{R}|H)},\label{eq:probOdds}\end{eqnarray}
where $\Theta(H)=p(\bar{H})/p(H)$ is the prior odds ratio for our
hypothesis. The prior odds ratio represents a likelihood ratio of the probability of 
$\bar{H}$ to the probability of $H$ and can be interpreted as our bias toward a 
galaxy teeming with intelligence (low prior odds) or a galaxy in which life is rare 
(high prior odds). Thus we see that our expression for $p(\bar{H}|\bar{V}_{R})$
depends on both the prior odds and the conditional probability $p(\bar{V}_{R}|H)$.

The term $p(\bar{V}_{R}|H)$ represents the probability that a search $\bar{V}_{R}$
returns null results when in fact NTAs actually do exist within $V$.
One case in which this may occur is if the spatial resolution $R$
(with dimensions of length) in the search $\bar{V}_{R}$ is too coarse to
resolve an NTA of diameter $d$; that is, if $R>d$. Here we are guaranteed
to miss any NTAs because their small size eludes observation. On the
other hand, if we somehow manage to completely survey our entire volume
$V$ so that $\bar{V}_{R}/V=1$, then we are guaranteed to find any NTAs
that are present so long as $R\le d$. We can use these limiting cases to write an equation for $p(\bar{V}_{R}|H)$ as 

\begin{equation}
\label{eq:condProb}
p(\bar{V}_{R}|H)=
\begin{cases}
1-\bar{V}_{R}/V & \text{for }R\le d\\
1.0 & \text{for }R>d
\end{cases}.
\end{equation} This allows us to express our conditional probability $p(\bar{V}_{R}|H)$
in terms of the relative volume searched $\bar{V}_{R}/V$ and the relative
size of our search resolution $R$ to the NTA diameter $d$.

\subsection{Equal Prior Odds}

We can simplify our analysis by considering the case of equal prior
odds such that $\Theta(H)=1$. We do not know values for the probabilities
$p(H)$ and $p(\bar{H})$, so perhaps it is reasonable to assume that
$p(H)=p(\bar{H})=0.5$. In other words, it is just as likely as not
that NTAs are present within a volume. With this simplifying approximation,
Eq. (\ref{eq:probOdds}) reduces to \begin{equation}\label{eq:probOdds2}
p(\bar{H}|\bar{V}_{R})=\frac{1}{1+p(\bar{V}_{R}|H)},
\end{equation} where we have used $p(\bar{V}_{R}|\bar{H})=1$, which is true by definition
(because a null search is guaranteed if there are no NTAs within the volume). Combining Eqs. (\ref{eq:probOdds2}) 
and (\ref{eq:condProb}), we find that\begin{equation}
p(\bar{H}|\bar{V}_{R})=\begin{cases} \frac{1}{2-\bar{V}_{R}/V} & \text{for }R\le d\\
0.5 & \text{for }R>d\end{cases}.\label{eq:equalPriorOdds}\end{equation}
In the case of equal prior odds, we find that our search-based confidence
in the non-existence of NTAs is related to the ratio of sampled volume
$\bar{V}_{R}$ at sufficient resolution $R\le d$ to the total volume $V$.
Thus, as we improve our search by increasing the amount of sampled
volume $\bar{V}_{R}$, we will steadily increase the likelihood $p(\bar{H}|\bar{V}_{R})$
that there are no NTAs within $V$ --- that is, unless we happen to
discover NTAs in the process.

\subsection{Non-Equal Prior Odds}

\begin{figure}
\includegraphics[width=5.5in]{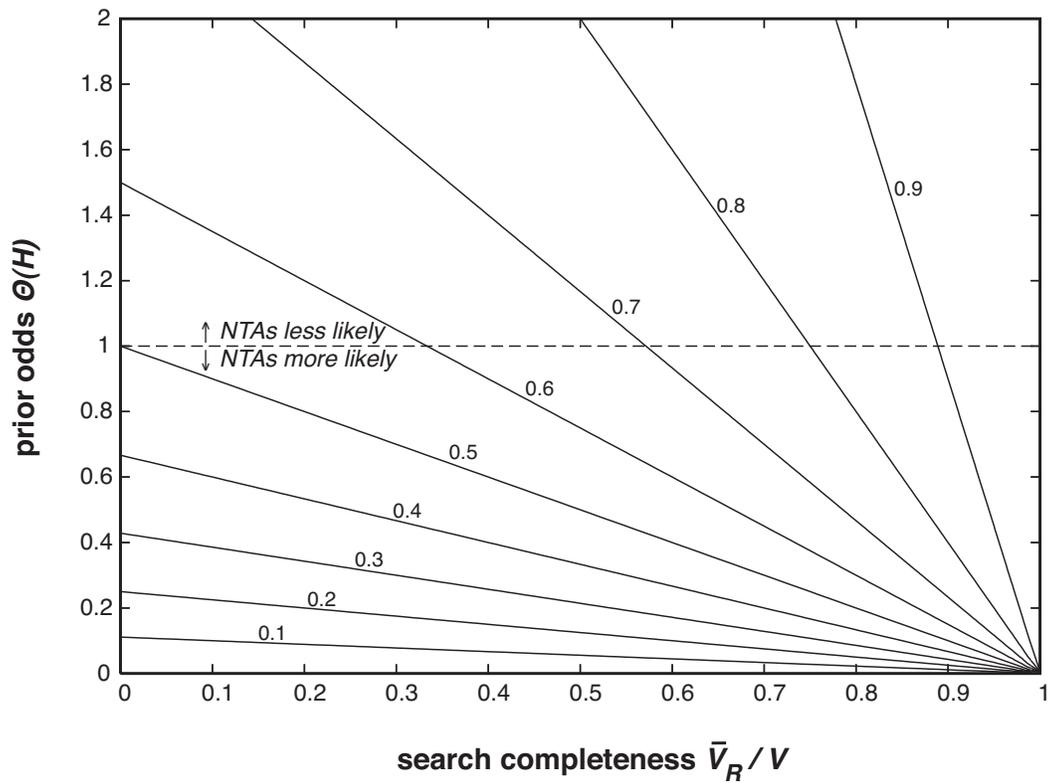}

\caption{The probability $p(\bar{H}|\bar{V}_{R})$ that NTAs are absent within a
volume as a function of prior odds $\Theta(H)$ and the search completeness
ratio $\bar{V}_{R}/V$. The dashed horiziontal line shows the case of equal
prior odds $\Theta(H)=1$. }\label{fig:prob2d}

\end{figure}

We can also consider the more general case where $\Theta(H)\ne1$.
Non-equal prior odds means that we have some reason to prefer one
of our hypotheses over the other; that is, there is a bias one way
or the other that NTAs are present within a volume. Substitution of
(\ref{eq:condProb}) into (\ref{eq:probOdds}) gives us \begin{equation}
p(\bar{H}|\bar{V}_{R})=\begin{cases}1-\frac{1-\bar{V}_{R}/V}{\Theta(H)+1-\bar{V}_{R}/V} & \text{for }R\le d\\
1-\frac{1}{\Theta(H)+1} & \text{for }R>d\end{cases}.\end{equation} This relationship is more complex than our assumption of equal prior
odds but allows us to consider cases where $p(H)>p(\bar{H})$, so
that $\Theta(H)<1$ (i.e. the probability of NTAs is high), or $p(H)<p(\bar{H})$,
so that $\Theta(H)>1$ (i.e. the probability of NTAs is low). We calculate
this likelihood $p(\bar{H}|\bar{V}_{R})$ as a function of both $\Theta(H)$
and the ratio $\bar{V}_{R}/V$ and show the results in Fig. \ref{fig:prob2d}.

Along the dashed horizontal line in Fig. \ref{fig:prob2d} where $\Theta(H)=1$,
we recover the expression in Eq. \ref{eq:equalPriorOdds} so that
$p(\bar{H}|\bar{V}_{R})=0.5$ at $\bar{V}_{R}/V=0$ with a steady increase as
the search $\bar{V}_{R}/V$ improves. Below this line, as prior odds $\Theta(H)$
decreases, we see that the likelihood $p(\bar{H}|\bar{V}_{R})$ also decreases.
This behavior is expected because low prior odds implies a high probability
of NTAs in the Solar System. Similarly, the likelihood $p(\bar{H}|\bar{V}_{R})$
increases above the dashed line as prior odds $\Theta(H)$ increases
becuase higher prior odds correspond to a low probability of NTAs.
This likelihood $p(\bar{H}|\bar{V}_{R})$ also increases as the search ratio
$\bar{V}_{R}/V$ increases, which again is consistent with our expectations
that a more comprehensive search should increase our confidence that
the Solar System is absent of NTAs.

\section{Discussion}

Our probabilistic analysis illustrates the difficulty in asserting
that the Solar System is absent of NTAs. Indeed, to search the entire
Solar System at the required spatial resolution to discover a 1 to
10 meter probe would be a monumental task, analogous to finding a
needle in a thousand-ton haystack (which is enough to cover an American
football field in a layer three feet deep). However, our volume $V$ need not be restricted to
the entire Solar System and can represent any planet or volume in
space. This allows us to use our probabilistic method to estimate
our certainty that systems such as Earth, the moon, or Mars are devoid
of NTAs.

The surface of Earth is one of the few places in the Solar System
that has been almost completely examined at a spatial resolution of less than 
1 meter. This is mostly because humans have spread across most of
the surface of Earth and would therefore have noticed any NTAs that
might exist. (Remaining consistent with the testable zoo hypothesis
discussed above, we do not consider NTAs that have been camouflaged
so that their detection is nearly impossible.) It may be a stretch
to claim that the entire surface of Earth, including land and oceans,
has really been searched at a sufficient resolution to discover NTAs;
after all, certain caves, jungles, and deserts as well as the complete
surface of the ocean may not have been scoured well enough. Additionally,
the subsurface of Earth has not been explored anywhere near the resolution
to discover such NTAs, if in fact they could reside in the deep ocean
or underground. However, nearby objects as small as 1 meter with Earth-bound trajectories should be detectable by orbital debris tracking systems. For the Earth volume, we can consider our search ratio
as being in the range $V/\bar{V}_{R}\approx0.7\text{ to }1.0$, which according
to Fig. \ref{fig:prob2d} may correspond to a probability $p(\bar{H}|\bar{V}_{R})\gtrsim0.75$ 
that Earth contains no NTAs --- at least where
prior odds $\Theta(H)\approx1$. This may be somewhat less than total certainty, but 
Earth is still the best-searched volume in the Solar System and gives us the highest confidence that
NTAs have not yet arrived.

One possible location for NTAs is the surface of the moon \cite{arkhipov1996,arkhipov1998}. The surface of the moon has been partially explored by human astronauts and lunar landers at a sufficient resolution to discover
any NTAs, but this in-situ survey is nowhere near complete. Other parts of
the lunar surface have been mapped by orbiting satellites, and the Lunar 
Reconnaissance Orbiter (LRO) is currently mapping the surface of the moon at a resolution 
as low as 0.5 meters \cite{chin2007,robinson2010}. We can
therefore consider our search ratio for the moon as $V/\bar{V}_{R}\lesssim0.5$,
which corresponds to a probability $p(\bar{H}|\bar{V}_{R})\lesssim0.65$ 
that the moon is absent of NTAs. As the LRO mission continues, surface 
maps of the moon will improve in resolution and further increase this probability. It is not 
necessarily obvious, though, that a high-resolution image of the lunar surface will 
automatically reveal any NTAs present; after all, a 10 meter satellite and a 10 meter boulder 
would appear similar in an LRO image. However, it has been suggested that NTAs might be identified 
by their emission of anomalous microwave or infrared radiation \cite{cornet2003}, so the discovery 
of an unexpected temperature anomaly on the lunar surface could be a signature worthy of further investigation. 
Likewise, the chemical composition of an NTA would presumably be different from the 
surrounding lunar regolith, so an anomalous spectral signature on the lunar surface 
could also be a potential indicator of an NTA. This suggests a method by which surface temperatures 
and mineralogies, such as those available from the LRO Diviner Lunar Radiometer 
Experiment \cite{glotch2010,greenhagen2010,hayne2010}, could be examined 
for anomalies that are consistent with the presence of an NTA. Such an analysis is 
worthy of future investigation but is beyond the scope of the present study.

Portions of the surface of Mars have been explored by the Mars Exploration Rover (MER) 
missions and other Martian landers at a sufficient resolution, but much less so than the moon. 
Mars-orbiting satellites likewise have improved our mapping of the planet's surface, but 
this is only enough to slightly raise our confidence that Mars is absent of NTAs. We can
consider our search ratio for Mars as $V/\bar{V}_{R}\lesssim0.1$,
which corresponds to a probability $p(\bar{H}|\bar{V}_{R})$ of about $0.5$
or lower that Mars is absent of NTAs. Continued Mars 
exploration by landers and rovers, as well as increased resolution on Mars-orbiting satellites, will 
slowly increase our understanding of the features of Mars' surface and will eventually permit 
a more quantitative analysis to search for spectroscopic or thermal anomalies as described above.

Other possible locations for NTAs within the Solar System include
the Earth-moon Lagrange points, the asteroid belt, and the Kuiper
belt. The five Lagrange points between any two orbiting bodies describe equilibrium gravitational positions where a satellite could in principle maintain an orbit without additional energy expenditure. However, radiation pressure would destabilize an orbit for a 1 to 10 meter object, so any active NTA would require some propulsion system in order to maintain a Lagrange point orbit. The asteroid belt between Mars and Jupiter also provides numerous hiding places and mineral resources for NTAs, while the more distant Kuiper belt offers a sparser distribution of rocky and icy bodies. At increasing distances from the Sun, none of these regions
have been searched well enough to discover NTAs. In these cases, our
search ratio would be at most $V/\bar{V}_{R}\lesssim10^{-3}$ or lower so
that we have almost no confidence that NTAs are absent in the Solar
System. Indeed, this point resonates with the basic message of our
argument: searches to date of the vast Solar System are sufficiently
incomplete that we cannot rule out the possibility that NTAs are present
and may even be observing us.

\section{Conclusion}

If other technological civilizations do in fact exist in the Milky
Way, then we must acknowledge the possibility that they may have chosen
to remotely explore our Solar System using unpiloted probes. Such
probes are being considered for human exploration of nearby star systems
and would have a limiting size of 1 to 10 meters. Although we may
be able to rule out the surface of Earth as a site that currently
harbors these NTAs, the rest of the Solar System has not been explored
at a sufficient resolution to discover these probes, if in fact they
do exist. Continued exploration of the lunar and Martian surfaces
will slowly increase our confidence in the absence of NTAs, as will
other Solar System missions that closely examine planetary bodies
(e.g. the Huygens probe landing on Titan), gravitationally stable Lagrange points \cite{freitas1980,valdes1983}, 
or asteroid debris belts \cite{forgan2011b}. Analysis of data from orbiting satellites, such as the LRO, may also help to constrain this probability by searching for thermal or compositional anomalies 
that are consistent with the presence of an NTA. Nevertheless,
the vastness of space implies that it will take some time before even
nearby objects can be conclusively considered devoid of NTAs.

In presenting this analysis, we do not intend to argue that the search
for NTAs in our Solar System should be given any sort of preference
in favor of other, more conventional astronomical missions. Our intention
is to present a framework by which we can estimate the completeness
of our search for NTAs, which will inevitably increase as we continue
to explore the moon, Mars, and other nearby regions of space. The
discovery of extraterrestrial technology would certainly be one of
the most significant findings in human history; for even if this technology
were non-functional, it would give us some certainty that life ---
and intelligence --- has developed elsewhere. With so many places
for a such small observational probe to hide in our own backyard,
we may as well keep our eyes open.

\section*{Acknowledgments} 
We thank Michael Busch, Seth Baum, and an anonymous reviewer for helpful suggestions that strengthened this paper. No 
funding was used for this research.





\bibliographystyle{model1-num-names}
\bibliography{haqqmisra_kopparapu}







\end{document}